\documentclass[conference]{IEEEtran}
\IEEEoverridecommandlockouts

\usepackage[letterpaper, left=0.65in,right=0.65in, top=0.75in,bottom=1in]{geometry}

\usepackage{amsmath,amsfonts}
\usepackage{amssymb}
\usepackage{amsthm}
\usepackage{array}
\usepackage[caption=false,font=normalsize,labelfont=sf,textfont=sf]{subfig}
\usepackage{textcomp}
\usepackage{stfloats}
\usepackage{url}
\usepackage{verbatim}
\usepackage{graphicx}
\usepackage{algorithm,algpseudocode}
\usepackage{cite}
\usepackage{bm}
\usepackage{arydshln}

\def\BibTeX{{\rm B\kern-.05em{\sc i\kern-.025em b}\kern-.08em
    T\kern-.1667em\lower.7ex\hbox{E}\kern-.125emX}}
\usepackage{balance}

\usepackage{xcolor}
\usepackage{array}

\newtheorem{theorem}{Theorem}

\newtheorem{example}{Example}

\newtheorem{corollary}{Corollary}

\floatname{algorithm}{Algorithm}

\usepackage{pgfplots}
\pgfplotsset{compat=newest}
\usetikzlibrary{plotmarks}
\usepackage{grffile}

\begin{document}
\title{On Transversality Across Two Distinct Quantum Error Correction Codes For Quantum Repeaters}
\author{
\IEEEauthorblockN{Mahdi Bayanifar$^1$, Alexei Ashikhmin$^2$, Dawei~Jiao$^1$, Olav Tirkkonen$^1$} 
             \\
        \IEEEauthorblockA{\em$^1$Department of Communications and Networking, Aalto University, Finland\\
              $^{2}$Nokia Bell Labs, Murray Hill, New Jersey, USA
              \\
              Email: \{mahdi.bayanifar, dawei.jiao,  olav.tirkkonen\}@aalto.fi} alexei.ashikhmin@nokia-bell-labs.com
     }
\maketitle
\begin{abstract}
In this paper, we investigate the transversality of pairs of CSS codes and their use in the second generation of quantum repeaters (QR)s. We show that  different stations of quantum link can experience different errors. Considering this fact, we suggest to use different CSS codes in different stations. 
We also suggest to use $[[n,k]]$ codes with $k>1$ as they are more efficient then codes with $k=1$. 
We establish sufficient and necessary conditions for a pair of CSS codes to be non-local CNOT-transversal. We show that in contrast to the well known CNOT transversality which states that two CSS codes should be the same, less restrictive constraints are needed.
Next, we establish sufficient and necessary conditions for a code pair to be CZ-transversal. 
\end{abstract}


\section{Introduction}
\label{introSec}

\IEEEPARstart{T}{he} 
 Quantum repeaters (QRs) are wildly used in quantum communication realm, as they can extent the distance and reliability of information transmission. Based on different protocols, there are three different generations of QRs\cite{jiang20092GQR}. 
In the first generation communication is established between different stations using shared entangled qubits. In order to improve the performance of these QRs, the second generation QRs has been introduced. Currently they  are most likely to be implemented in near-term quantum devices \cite{van20202Grencent}. These QRs rely on using quantum error correction codes (QECCs) 
for achiving high fidelity entangled pairs. In the third generation of QRs, information is encoded with QECCs and directly transmitted between nearby stations without using shared entangled pairs.

QECCs encode into logical qubits into physical qubits and protect the logical information from quantum noises. In the second generation QRs, QECCs are used in each station. The communication protocol is based on logical entanglement between different stations, and in order to avoid  error propagation the protocol should be fault-tolerant, which requires that QECCs used in neighboring stations were CNOT-transversal \cite{Gottesman:2022jlv}.

The most popular QECCs are the stabilizer codes where code vectors are stabilized by operators forming a commutative group \cite{gottesman1997stabilizer}. The CSS codes are a special case of the stabilizer codes in which the generators of the commutative group are pure $X$ and $Z$ Pauli operators \cite{nielsen00}. 
In \cite{shor1996transversal} it is shown that any CSS code is Pauli and CNOT-transversal and any self-orthogonal CSS code is Hadamard and CZ-transversal.

In our previous work \cite{DaweConfPaper23}, we analyze error models appearing in the communication protocol of the second generation QRs. We showed that the errors occurring in neighboring  stations are biased and correlated. For instance, the Pauli $X$ error could be more likely in one statition and $Z$ errors more likely in the neighboring station. 
The reason for this bias arises from the Bell state purification and remote CNOT procedures \cite{zhou2000}. 

This motivates us to design and optimize QECCs specifically for the above error modes. Intuitively, we want to  
use different CSS codes in nearby stations, one with larger resistance to $X$ errors, and another with larger resistance to  $Z$ errors \cite{DaweConfPaper23}. However, such CSS codes may be non transversal, which will lead to an nont fault-tolerant communication protocol. This motivates us to study conditions on code pairs be transversal.  

It is worth noting that the transversality between two different codes is not only used for QRs system, but also can be applied in quantum computation, like distributed quantum computation, or other realms. The no-go theorem considers transversality  and computational universality under the assumption of using same code in every code block. 
 Our approach establishes the transversality between different codes and may provide a new path to the fault-tolerant universal quantum computation.

Motivated by the above reasons, we study this paper the non-local CNOT and CZ-transversality of pairs of CSS codes.  
In our studies, we first rovide conditions on the non-local CNOT-transversality between CSS codes used in nearby stations. We observe that in contrast to the well known fact that for having a CNOT transversal gate, one needs to have the same code in the station, less restrictive conditions is needed. Then, we investigate the transversality of the non-local CZ gates and find sufficient conditions for achieving the CZ-transversality. As an example of transversal CZ-gate, we investigate the special structure which is considered in our previous work \cite{DaweConfPaper23}, called mirrored structure. This structure could get better result under the second generation QRs error model than using the same code in every station. We show that any mirrored structure could achieve sufficient condition of non-local CZ-transversality. Finally, through some examples, we show that for achieving the transversality, one needs to select the mapping properly and otherwise the transversality may not hold. 


The paper is organized as follows: We review basic definitions of CSS codes in ~\ref{Sec:PrelSec}. In 
Section \ref{Sec:QRs} we consider communication protocol and error models. Next, we establish conditions on code pairs to be CNOT and CZ-transversal~\ref{Sec:Transvers}. 
Effects of the encoding mappings are discussed in Section~\ref{Sec:MappingEffect}, and  Section~\ref{Sec:Concl} concludes the paper. 

\section{Qubits, Quantum Operations, and CSS codes}
\label{Sec:PrelSec}
In this section, we recall the main definitions of quantum CSS codes. More details on this can be found, e.g., in 
\cite{nielsen00}. 

Let $\mathbf{v} = \left( v_1, ..., v_n\right) \in \mathbb{F}_2^n$ and $|0\rangle = (1,0)^T,\ |1\rangle =(0,1)^T \in \mathbb{C}^2$.  Then the quantum states 
$|\mathbf{v}\rangle = |v_1\rangle \otimes \ldots |v_n\rangle$ 
form the computation basis of $\mathbb{C}^{2^n}$, and any pure state $|\psi\rangle \in \mathbb{C}^{2^n}$ of $n$ qubits can be written in the form 
\begin{align}
|\bm{\psi}\rangle = 
&\sum_{\mathbf{v}\in \mathbb{F}_2^n } 
\alpha_{\mathbf{v}} |\mathbf{v}\rangle, \\
\mbox{where } &\sum_{\mathbf{v}\in \mathbb{F}_2^n } |\alpha_v|^2 = 1.\nonumber 
\end{align}
The CNOT gate between a control qubit in a pure state $|\psi\rangle$ and target qubit in $|\xi\rangle$ corresponds to the unitary transformation 
$\mathbf{U}_{CNOT}({\lvert\psi\rangle\otimes |\xi\rangle})$, where 
$$
\mathbf{U}_{CNOT } =  \begin{bmatrix}
        1 & 0 & 0 & 0  \\
        0 & 1 & 0 & 0\\
        0 & 0 & 0 & 1\\
        0 & 0 & 1 & 0
    \end{bmatrix}. 
$$
For $a,b\in \mathbb{F}_2$
we have $\mathbf{U}_{ CNOT } \left( |a\rangle \otimes |b\rangle \right) = |a\rangle \otimes |a+b\rangle$. Denote by $\mathbf{U}_{CNOT,i,i+n}\in \mathbb{C}^{2^{2n}}$ the gate that conducts the CNOT for qubits $i$ and $n+i$ and leave other qubits untouched. 
Assume that we have two sets of $n$ qubits in pure states 
$$
|\bm{\psi} \rangle  = \sum_{\mathbf{v}\in \mathbb{F}_2^n } 
\alpha_{\mathbf{v}} |\mathbf{v}\rangle,\mbox{ and } |\bm{\xi}\rangle = \sum_{\mathbf{w}\in \mathbb{F}_2^n } 
\beta_{\mathbf{w}} |\mathbf{w}\rangle,  
$$
and denote by $CNOT$ the operator that conducts CNOT gates for all the qubit pairs $(i,n+i),i=1,\ldots,n$. It is not difficult to see that 
\begin{align}
&CNOT( |\bm{\psi}\rangle \otimes |\bm{\xi}\rangle)\nonumber \\
= &\left( \mathbf{U}_{CNOT,1,n+1}\mathbf{U}_{CNOT,2,n+2} \ldots \mathbf{U}_{CNOT,n,2n} \right) \left( |\bm{\psi}\rangle \otimes |\bm{\xi}\rangle\right) \nonumber\\
=&\sum_{\mathbf{v}\in \mathbb{F}_2^n } \sum_{\mathbf{w}\in \mathbb{F}_2^n } 
\alpha_{\mathbf{v}} \beta_{\mathbf{w}} |\mathbf{v} \rangle\otimes |\mathbf{v} + \mathbf{w}\rangle. \label{eq:CNOT as sum}
\end{align}
Similar definitions can be made for quantum mixed states, but we omit those details. 
Recall that Control-Z (CZ) gate is defined by $\mathbf{U}_{CZ} =\mbox{diag}(1,1,1,-1)$. We denote by $CZ$ the operator that applies CZ gates to all the qubit pairs $(i,n+i),i=1,\ldots,n$. 
It is again not difficult to see that
\begin{equation}
 CZ( |\bm{\psi}\rangle \otimes |\bm{\xi} \rangle)
= \sum_{\mathbf{v}\in \mathbb{F}_2^n } 
\sum_{\mathbf{w}\in \mathbb{F}_2^n } (-1)^{\mathbf{v}
\mathbf{w}^T}
\alpha_{\mathbf{v}} \beta_{\mathbf{w}} 
|\mathbf{v} \rangle
\otimes |\mathbf{w} \rangle. \label{eq:CZ as sum}
\end{equation}

The widely used completely depolarizing error model is described by the Pauli matrices:  
\begin{equation}
    \mathbf{X} \triangleq
    \begin{bmatrix}
        0 & 1 \\
        1 & 0
    \end{bmatrix},
    \;
    \mathbf{Z} \triangleq
    \begin{bmatrix}
        1 & 0 \\
        0 & -1
    \end{bmatrix}
    \;
    \mathbf{Y} \triangleq
    \begin{bmatrix}
        0 & -i \\
        i & 0
    \end{bmatrix},
\end{equation}
where $i\triangleq \sqrt{-1}$. For $\mathbf{a},\ \mathbf{b}\in \mathbb{F}_2^n$ we define operator 
$$
\mathbf{D}\left( \mathbf{a, b} \right) = \mathbf{X}^{a_1}\mathbf{Z}^{b_1} \otimes ... \otimes \mathbf{X}^{a_n}\mathbf{Z}^{b_n}.
$$ 

Let $\mathcal{C}_1$ and $\mathcal{C}_2$ be two classical linear codes with parameters $[n, k_1, d_1]$ and $[n, k_2, d_2]$, respectively. By $\mathcal{C}_1^\perp$ and 
$\mathcal{C}_2^\perp$ we denote their dual codes of dimensions $k_1^\perp$ and $k_2^\perp$ respectively.  
Let also the property $\mathcal{C}_2^{\perp} \subset \mathcal{C}_1$ hold. Then $\mathcal{C}_1$ and $\mathcal{C}_2$ define  an $[[n, k, d]], k=k_1+k_2-n, d = \min\left(d_1, d_2 \right)$, quantum $\text{CSS}\left(\mathcal{C}_1, \mathcal{C}_2\right)$ code, which is a linear subspace of dimension $2^k$ in $\mathbb{C}^{2^n}$. 
Let us assume that there is a linear bijection between vectors
$\bm{\psi}\in \mathbb{F}_2^k$ and representatives 
$\mathbf{x}\in \mathcal{C}_1$
of cosets of $\mathcal{C}_2^\perp$ in the quotient group $\mathcal{C}_1/\mathcal{C}_2^\perp$. Then $\text{CSS}\left(\mathcal{C}_1, \mathcal{C}_2\right)$ poses the orthogonal basis 
\begin{equation}
	| \bm{\psi} \rangle_L = \frac{1}{\sqrt{\left\lvert \mathcal{C}_2^{\perp}\right\rvert} }\sum_{\mathbf{y} \in \mathcal{C}_2^{\perp}}{| \mathbf{x}+\mathbf{y} \rangle}. \label{Eq:StablStCSS}
\end{equation}
 
It is worth noting that the CSS codes are special case of the stabilizer codes.
This means that for any $[[n,k,d]]$ CSS code $Q$, we can find a commutative group $\mathcal{S}, |\mathcal{S}| = 2^{n-k}$, composed by operators of the form $\gamma_{\mathbf{a},\mathbf{b}} D({\bf a},{\bf b})$, and for given $(\mathbf{a},\mathbf{b})$ we have $\gamma_{\mathbf{a},\mathbf{b}}$ is either $1$ or $-1$.
For any $\gamma_{\mathbf{a},\mathbf{b}} D({\bf a},{\bf b})\in \mathcal{S}$ we have that 
$$ \gamma_{\mathbf{a},\mathbf{b}} D({\bf a},{\bf b}) 
|\bm{\psi}\rangle = |\bm{\psi}\rangle, \mbox{ for any } |\bm{\psi}\rangle\in Q. 
$$  
The vectors $({\bf a},{\bf b})$, 
defining the operators $\gamma_{\mathbf{a},\mathbf{b}} D({\bf a},{\bf b}) \in \mathcal{S}$, form the linear code with the generator matrix 
$$
\mathbf{G}^{\mathcal{Q}} = 
    \left[
    \begin{array}{c;{2pt/2pt}c}
        \mathbf{G}_2^\perp & \mathbf{0} \\
        \hdashline
        \mathbf{0} & \mathbf{G}_1^\perp 
    \end{array}
    \right]. 
$$
\section{Quantum Network with Quantum Codes Matched to Error Model}
\label{Sec:QRs}

\begin{figure}
    \centering
    \includegraphics[scale=0.35]{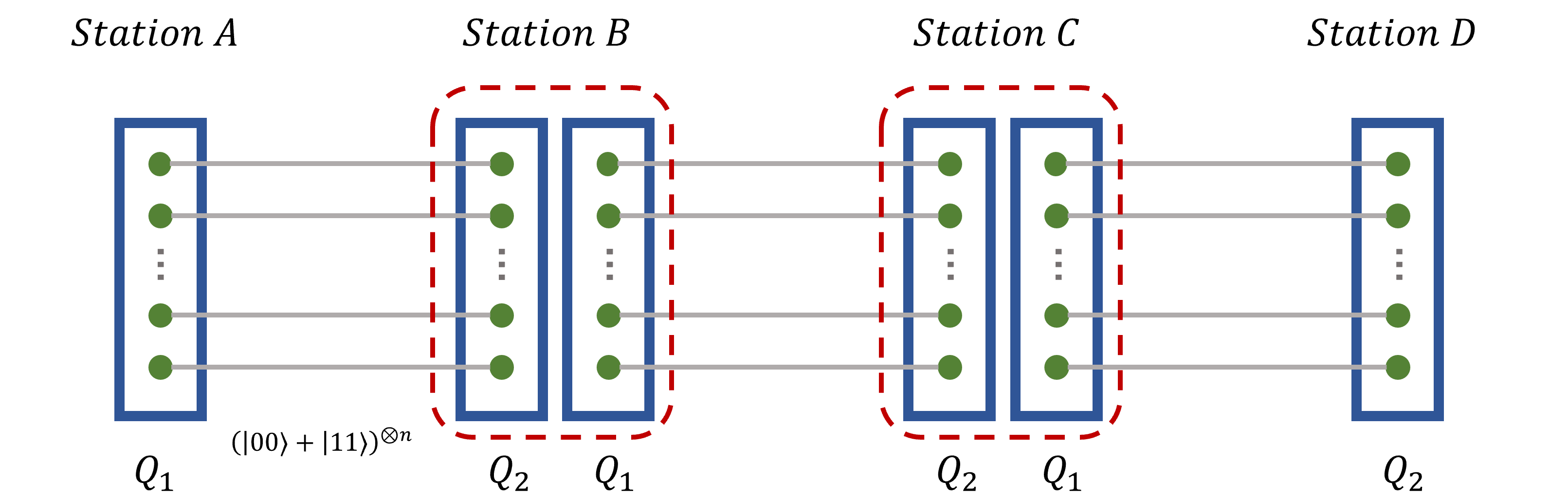}
    \caption{$Q_1$ and $Q_2$ are different CSS codes with higher capability for correcting Pauli $Z$ and $X$ errors. These codes are arranged in such way and entangle by Bell states. }
    \label{fig:2GQR}
\end{figure}

A typical quantum network link is shown in Fig.\ref{fig:2GQR}. The intermediate stations contain QRs and possibly other hardware. In this work we assume that QRs of the second generation are used.  Compared with other generations these QRs look  mostly promising as they could be implemented with near-term devices. The second generation QRs use quantum codes to suppress the procedure errors, as we discuss it below. Using quantum codes is efficient in terms of achieving high fidelity and makes the requirements on the quantum hardware and its control relevantly low and therefore more achievable with near-term quantum devices \cite{van20202Grencent}.

The communication protocol of the second generation QRs typically assumes that 
the same $[[n,1]]$ CSS code is used in all quantum stations along a communication link. However, from the information-theoretic point of view it is more efficient to use $[[n,k]]$ codes with $k>1$ rather then repetitive use of $[[n/k,1]]$ codes. Thus, following our recent work \cite{DaweConfPaper23}, 
we assume CSS codes with $k>1$. 

Next, as we argue below, type of errors in stations along a quantum link can vary significantly. 
So, we suggest to use different CSS codes in different stations, and match the codes to  particular error models of the stations. We assume that the neighboring 
stations A and B use 
$CSS(\mathcal{C}_1,\mathcal{C}_2)$, 
and $CSS((\mathcal{C}_3,\mathcal{C}_4)$ 
codes respectively, where $\mathcal{C}_3$ and $\mathcal{C}_4$, $\mathcal{C}_4^{\perp} \subset \mathcal{C}_3$, are classical codes  
with parameters $[n, k_3, d_3]$ and $[n, k_4, d_4]$ respectively. 
By $\mathcal{C}_3^\perp$ and 
$\mathcal{C}_4^\perp$ we denote their dual codes of dimensions $k_3^\perp$ and $k_4^\perp$ respectively. 
We assume that the following protocol is conducted between neighboring stations. 

{\bf Local Swapping Protocol} 

1. Stations A and B prepare their $k$ logical qubits $q_1^A, \ldots,q_k^A$ and 
$q_1^B,\ldots,q_k^B$ in the states $q_j^A=\lvert+\rangle_L=\lvert0\rangle_L+\lvert1\rangle_L$, and $q_j^B=\lvert0\rangle_L$, $j=1,\ldots,k$, respectively. Further they encode the logical qubits into $n$ physical qubits
$p_1^A, \ldots ,p_n^A$ and $p_1^B, \ldots ,p_n^B$
with  $CSS(\mathcal{C}_1,\mathcal{C}_2)$ and  
$CSS(\mathcal{C}_3,\mathcal{C}_4$) codes respectively.  

2. Station A generates $N>n$ Bell pairs qubits, e.g, photons, in $\lvert00\rangle+\lvert11\rangle$ state, and sends the second qubit of each pair to station B via a classical link, e.g., optic fiber.

3. Stations A and B conduct the purification procedure for the $N$ Bell pairs, see \cite{terhal2002entanglement}, and obtain $n$ shared (noisy) Bell pairs $c_1^A, c_1^B; \ldots ;c_n^A, c_n^B$. 

\begin{figure}
    \centering
    \includegraphics[scale=0.7]{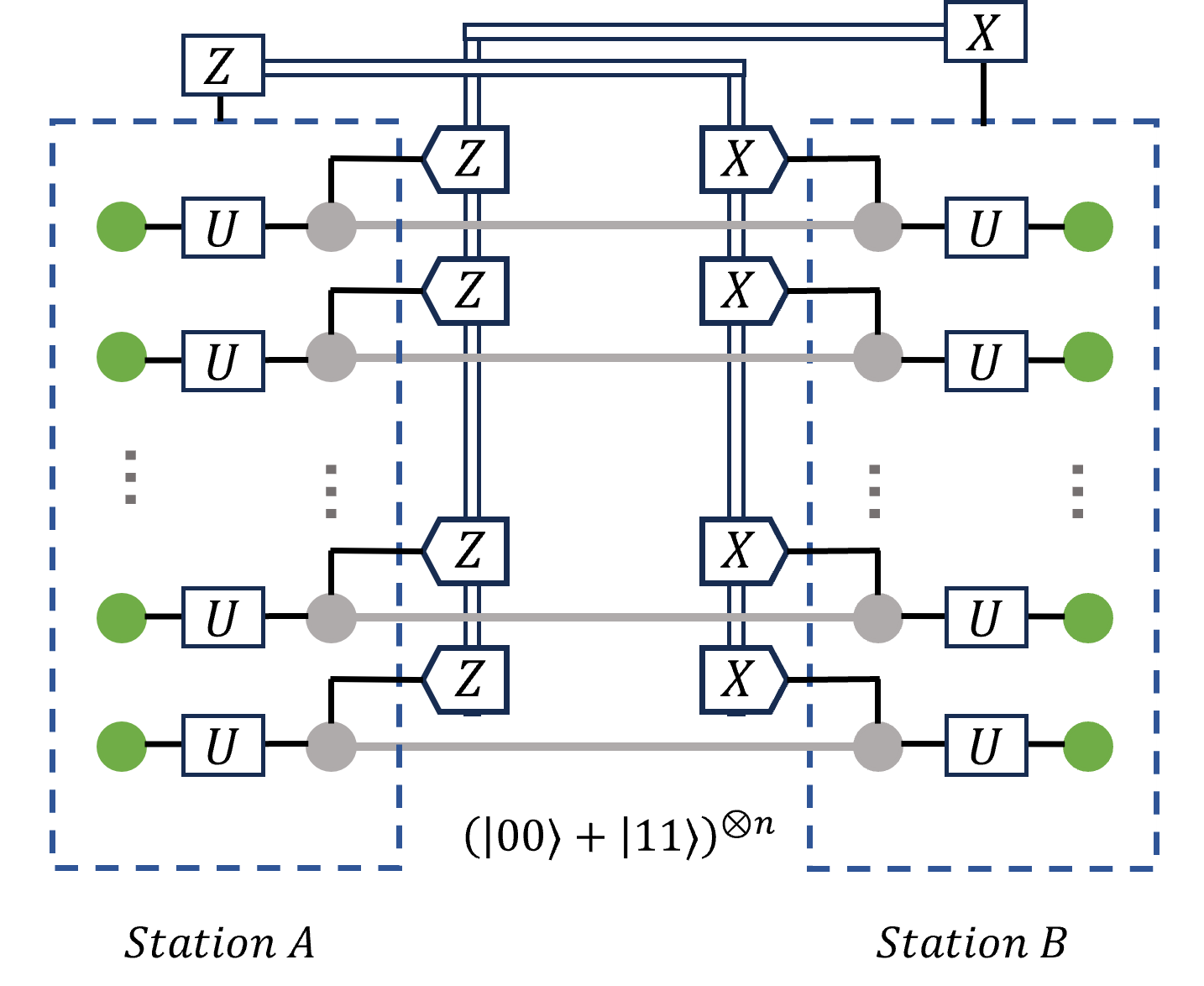}
    \caption{The transversal non-local gate schematic: Unitary operator $\mathbf{U}$ can be selected as CNOT or CZ gate in order to achieve transversal non-local CNOT (green qubit as control) or CZ gate. After implementing $\mathbf{U}$ operator, these Bell states (gray dots), are measured in Pauli $Z$ and $X$ bases. According to the outcome, the feedback Pauli $X$ and $Z$ operation will be applied on physical qubits (green dots) in each station. 
    }
    \label{fig:nonlocalgate}
\end{figure}

4. Stations A and B conduct local operations shown in Fig.\ref{fig:nonlocalgate} using their qubits 
$p_1^A, \ldots,p_n^A$ (green dots) and $\ c_1^A, \ldots,c_n^A$ (gray dots) and $p_1^B, \ldots,p_n^B,\ c_1^B, \ldots,c_n^B$ respectively, with $\mathbf{U} = \mathbf{U}_{CNOT}$. By doing this Stations A and B effectively conduct remote 
CNOT operations between $p_j^A$ and  $p_j^B$, $j=1,\ldots,n$.

5. Stations A and B correct errors in their physical qubits using decoders of $CSS(\mathcal{C}_1,\mathcal{C}_2)$ and $CSS(\mathcal{C}_3,\mathcal{C}_4)$. 

It is important to note that under the assumption that codes $CSS(\mathcal{C}_1,\mathcal{C}_2)$ and $CSS(\mathcal{C}_3,\mathcal{C}_4)$ are CNOT-transversal, at Step 5 the physical qubits at Station A and B are turned into the encoded 
states of these codes corresponding to logical qubits $|q_j^A\ q_j^B\rangle = |00\rangle+|11\rangle$, 
$j=1,\ldots,k$ (we discuss this in details below). Therefore conducting the above protocol for all the neighboring Stations, and further applying entanglement swapping one achieves long-distance entanglement between $k$ logical qubits \cite{bos98EntSwap}.
 

The reason for using CSS codes at Step 5 is that high  logical fidelity of entanglement can be achieved with relevantly low cost. For achieving the same fidelity with Bell state purification, the resource cost, such as qubit number and time cost, would be much larger  \cite{muralidharan2016optimal}. 


The purified Bell states at Step 3 are still noisy and that noise propagates to physical qubits $p_1^A, \ldots,p_n^A, p_1^B,\ldots,p_n^B$ through the gate of the circuit shown in Fig.\ref{fig:nonlocalgate}. This results in a complex error model for the physical qubits. One such model was introduced and analysed recently in \cite{DaweConfPaper23}. Let $\rho$ denote the error-free joint density matrix of physical qubits $p_j^A$ and $p_j^B$ for some fixed $j$ after conducting the remote CNOT. Then, according to the error model from  \cite{DaweConfPaper23}, the physical qubits will have the density matrix  
 \begin{eqnarray} 
      \mathcal{N}(\bm{\rho})&=& \left( 1-\sum_{i=1}^3 f_i \right)[\mathbf{I}_A \mathbf{I}_B](\bm{\rho})~+~f_1[\mathbf{Z}_A\mathbf{I}_A](\bm{\rho})  \cr
      &&~+~f_2[\mathbf{I}_A\mathbf{X}_B](\bm{\rho})~+~              f_3[\mathbf{Z}_A \mathbf{X}_B](\bm{\rho}),
        \label{Eq.Purified_Model}
\end{eqnarray}
where $[\mathbf{U}_1 \mathbf{U}_2](\bm{\rho})$ 
denotes 
$(\mathbf{U}_1 \otimes \mathbf{U}_2) \bm{\rho} (\mathbf{U}_2^\dag\otimes \mathbf{U}_1^\dag)$.  Note that $f_1$ is the probability of errors $Z$ in Station A and the absence of errors in Station B, while $f_2$ is the probability of $X$ errors in Station B, and $f_3$ is the probability of correlated errors. It was observed in \cite{DaweConfPaper23} that typically one type of errors  dominates. For example, we may have that $f_1>f_2>>f_3$. So, it is desirable to use different CSS codes, say $Q_1$ and $Q_2$ in Stations A and B, as it is shown in Fig.\ref{fig:2GQR}. By adjusting these codes to error models of Stations A and B, we can significantly improve the fidelity, reduce time cost, and so on. However, it is important to remember that we cannot use arbitrary CSS codes since the codes should be CNOT-transversal. 

In the next Sections we study the main principles of construction of CNOT-transversal CSS codes. 
  In  \cite{DaweConfPaper23} it was shown that if CZ-transversal codes are used in Stations A and B, then the 
  Local Swapping Protocol can be modified, with the help of using magic-state operation so that one can still implement the needed  remote CNOT operation. For this reason, we also study below construction of CZ-transversal codes. 

We also believe that pair-wise transversal codes will find other applications beyond their use in QRs, and so may constitute important research directions.   

\section{Transversality of CSS Codes}\label{Sec:Transvers}

 Firs we would like to recall the well known fact that if the same CSS code is used in Stations A and B then we have CNOT-transversality granted and therefore can implement Local Swapping Protocol. 
However, as we explained above, due to asymmetry of errors it would more beneficial to use different CSS codes in neighboring Stations. 
 

Let us assume that we have two $\text{CSS}(\mathcal{C}_1, \mathcal{C}_2)$ and $\text{CSS}(\mathcal{C}_3, \mathcal{C}_4)$ codes in stations A and B, respectively. Our objective is to find suitable conditions that would guarantee CNOT and/or CZ-transversality of these codes. Denote by $\mathbf{G}_2^{\perp}$ and $\mathbf{G}_4^{\perp}$ be generator matrices of $\mathcal{C}_2^{\perp}$ and $\mathcal{C}_4^{\perp}$, respectively. Then, the generator matrices of $\mathcal{C}_1$ and $\mathcal{C}_3$ can be written in the following form 
\begin{equation} \label{Eq:G1G2CSS}
    \mathbf{G}_1 = \begin{bmatrix}
        \mathbf{G}_2^{\perp} \\ 
        \mathbf{A}
    \end{bmatrix}, \qquad
    \mathbf{G}_3 = \begin{bmatrix}
        \mathbf{G}_4^{\perp} \\ 
        \mathbf{B}
    \end{bmatrix},
\end{equation}
where $\mathbf{A}$ and $\mathbf{B}$ are $k\times n$ binary matrices of rank $k$. 

We will use $\mathbf{A}$ and 
$\mathbf{B}$ for defining the linear bijections between vectors
$\bm{\psi}^A, \bm{\psi}^B \in \mathbb{F}_2^k$
and representatives 
$\mathbf{x}^A \in \mathcal{C}_1$ and $\mathbf{x}^B \in \mathcal{C}_3$ 
of cosets in the quotient groups $\mathcal{C}_1/\mathcal{C}_2^\perp$ and $\mathcal{C}_3/\mathcal{C}_4^\perp$. With these bijections the code vectors of $\text{CSS}\left( \mathcal{C}_1, \mathcal{C}_2\right)$ and $\text{CSS}\left( \mathcal{C}_3, \mathcal{C}_4\right)$  corresponding to logical qubits in the states $\vert \bm{\psi}^A \rangle$ and $\vert \bm{\psi}^B \rangle$ are 
\begin{align}
    \vert \bm{\psi}^A \rangle_L & = \frac{1}{\sqrt{\lvert C_2^{\perp} \rvert}}\sum_{\mathbf{y} \in C_2^{\perp}}{|\mathbf{x}^A + \mathbf{y} \rangle } \label{Eq:LogQuSA}   \\
    \vert \bm{\psi}^B \rangle_L & = \frac{1}{\sqrt{\lvert C_4^{\perp} \rvert}}\sum_{\mathbf{z} \in C_4^{\perp}}{|\mathbf{x}^B + \mathbf{z} \rangle },\label{Eq:LogQuSB}
\end{align}
where $\mathbf{x}^A = \bm{\psi}^A \mathbf{A}$ and $\mathbf{x}^B = \bm{\psi}^B \mathbf{B}$.

\begin{theorem}\label{thm:CNOTtransversality}
    Codes $\text{CSS}\left( \mathcal{C}_1, \mathcal{C}_2\right)$ and $\text{CSS}\left( \mathcal{C}_3, \mathcal{C}_4\right)$ are CNOT-transversal, with $\text{CSS}\left( \mathcal{C}_1, \mathcal{C}_2\right)$ being control and $\text{CSS}\left( \mathcal{C}_3, \mathcal{C}_4\right)$ target codes, 
    iff
    \begin{align}
        \mathcal{C}_2^\perp & \subseteq \mathcal{C}_4^\perp \label{Eq:CNOTcond1} \\
        \mathcal{C}_1/\mathcal{C}_2^{\perp} & \cong \mathcal{C}_3/\mathcal{C}_4^{\perp}\label{Eq:CNOTcond2}
    \end{align}
\end{theorem}
\begin{proof}
    For showing the transversality, we need to show that applying CNOT gates to $k$ pairs of logical qubits in the states $\vert \bm{\psi}^A \rangle$ and $\vert \bm{\psi}^B \rangle$, and then encoding the results into code vectors of $\text{CSS}(\mathcal{C}_1,\mathcal{C}_2)$ and $\text{CSS}(\mathcal{C}_3,\mathcal{C}_4)$ gives the same results as applying CNOT operation to $n$ pairs of physical qubits in  states $\vert \bm{\psi}^A \rangle_L$ and $\vert \bm{\psi}^B \rangle_L$ defined in  
    (\ref{Eq:LogQuSA}) and (\ref{Eq:LogQuSB}). 
    
     If we first apply CNOT gates to logical qubits and then conduct encoding into code vectors of $\text{CSS}(\mathcal{C}_1,\mathcal{C}_2)$ and $\text{CSS}(\mathcal{C}_3,\mathcal{C}_4)$. According to (\ref{eq:CNOT as sum}),   we will get the result
    \begin{equation} \label{Eq:LogCNOTDef}
     | \bm{\psi}^A  \rangle_L \otimes | \bm{\psi}^A  \oplus \bm{\psi}^B  \rangle_L,
    \end{equation}
    where using definition in \eqref{Eq:LogQuSA}, and according to \eqref{Eq:StablStCSS} and \eqref{Eq:G1G2CSS},
    \begin{equation}
       \vert \bm{\psi}^A \oplus  \bm{\psi}^B \rangle_L  = \frac{1}{\sqrt{\lvert \mathcal{C}_4^{\perp} \rvert}}\sum_{\mathbf{y} \in C_4^{\perp}}{|\left( \bm{\psi}^A  +\bm{\psi}^B\right) \mathbf{B}+\mathbf{y} \rangle }. \label{Eq:EqivalLogCnot}
    \end{equation} 
    Next, if we apply CNOT gates to the $n$ pairs of physical qubits encoded into states defined in  (\ref{Eq:LogQuSA}) and (\ref{Eq:LogQuSB}) then, according to (\ref{eq:CNOT as sum}), we get the state
    \begin{align}
	   & \text{CNOT} (| \bm{\psi}^A  \rangle_L \otimes | \bm{\psi}^B \rangle_L)  \nonumber \\
    &= \frac{1}{\sqrt{ \left\lvert \mathcal{C}_2^{\perp} \right\rvert \left\lvert \mathcal{C}_4^{\perp} \right\rvert }} \sum_{\mathbf{y} \in \mathcal{C}_2^{\perp}, \mathbf{z} \in \mathcal{C}_4^{\perp}}{ | \mathbf{x}^A + \mathbf{y} \rangle \otimes  | \mathbf{x}^A + \mathbf{x}^B + \mathbf{y+z} \rangle} \nonumber \\
        & \stackrel{(a)}{=} \frac{1}{\sqrt{ \left\lvert \mathcal{C}_2^{\perp} \right\rvert \left\lvert \mathcal{C}_4^{\perp} \right\rvert }} \sum_{\mathbf{y} \in \mathcal{C}_2^{\perp}, \mathbf{z}' \in \mathcal{C}_4^{\perp}}{ | \mathbf{x}^A + \mathbf{y} \rangle \otimes  | \mathbf{x}^A + \mathbf{x}^B + \mathbf{z}' \rangle}, \label{Eq:FinCNOT} 
\end{align}
where $(a)$ is true iff $\mathcal{C}_2^{\perp} \subseteq \mathcal{C}_4^{\perp}$ or equivalently $\mathcal{C}_4 \subseteq \mathcal{C}_2$. 
 
For achieving the transversality we need that \eqref{Eq:EqivalLogCnot} be equal to \eqref{Eq:FinCNOT} for any $\bm{\psi}^A$ and $\bm{\psi}^B$. This is possible if and only if $\mathbf{A=B}$, and this means that  
$\mathcal{C}_1/\mathcal{C}_2^{\perp} \cong \mathcal{C}_3/\mathcal{C}_4^{\perp}$. Note that the cosets in $\mathcal{C}_1/\mathcal{C}_2^{\perp}$ and $\mathcal{C}_3/\mathcal{C}_4^{\perp}$
may contain different number of vectors, since $\mathcal{C}_2^{\perp}$ can be smaller than $\mathcal{C}_4^{\perp}$, but the quotient groups are still isomorphic if 
$\mathbf{A=B}$. 
\end{proof}
Note that considering \eqref{Eq:G1G2CSS} and CNOT-transversality constraints given in \eqref{Eq:CNOTcond1} and \eqref{Eq:CNOTcond2}, we can rewrite the generators of $\mathcal{C}_1$ and $\mathcal{C}_2$ as follows
\begin{equation} \label{Eq:G1G2CNOTtrans}
    \mathbf{G}_1 = \begin{bmatrix}
        \mathbf{G}_2^{\perp} \\ 
        \mathbf{A}
    \end{bmatrix}, \qquad
    \mathbf{G}_3 = \begin{bmatrix}
        \mathbf{G}_2^{\perp} \\ 
        \mathbf{D} \\
        \mathbf{A}
    \end{bmatrix},
\end{equation}
where $\mathbf{D}$ is a $(k_4^\perp-k_2^\perp) \times n$ matrix of the rank $k_4^\perp-k_2^\perp$, which serves as a generator of $\mathcal{C}_4^\perp / \mathcal{
C}_2^\perp$.  Note that this structure implies that $\mathcal{C}_1 \subseteq \mathcal{C}_3$ and may prompt one to conclude that it is sufficient that $\mathcal{C}_1 \subset \mathcal{C}_3$ in order to have the CNOT-transversality. However, this is not correct. For example we can consider the case that $\mathcal{C}_1 = \mathcal{C}_4^\perp \subset \mathcal{C}_3$, i.e.,
\begin{equation*}
    \mathbf{G}_1 = \begin{bmatrix}
        \mathbf{G}_2^{\perp} \\ 
        \mathbf{A}
    \end{bmatrix}, \qquad
    \mathbf{G}_3 = \begin{bmatrix}
        \mathbf{G}_2^{\perp} \\ 
        \mathbf{A} \\
        \mathbf{\mathbf{D}}
    \end{bmatrix}.
\end{equation*}
In this example $\mathcal{C}_1 \subset \mathcal{C}_3$, however we can observe that this configuration does not satisfy the property given in \eqref{Eq:CNOTcond2}. Below we give an example of two different CSS codes that CNOT-transversal.

\begin{example} The following CSS codes of length $n=7$ with $k_1 = 4$, $k_2=5$, and 
$k_3 = 5$, $k_4=4$ respectively and generator matrices     
    \begin{align*}
       \mathbf{G}_1 &= 
       \begin{bmatrix}
           \mathbf{G}_2^\perp \\
           \mathbf{A}
       \end{bmatrix}
        =
        \left[
        \begin{array}{ccccccc}
            1&1&0&0&0&0&0 \\
            0&1&0&1&1&1&1\\
            \hdashline
            0&0&1&1&0&1&1 \\
            1&0&1&1&1&0&0
        \end{array}
        \right]\\
        \mathbf{G}_3 & =
    \begin{bmatrix}
           \mathbf{G}_2^\perp \\
           \mathbf{D} \\
           \mathbf{A}
    \end{bmatrix}
    =
    \left[ 
    \begin{array}{ccccccc}
        1&1&0&0&0&0&0 \\
        0&1&0&1&1&1&1 \\
        \hdashline
        0&1&1&1&0&1&0 \\
        \hdashline
        0&0&1&1&0&1&1 \\
        1&0&1&1&1&0&0
    \end{array} \right],
    \end{align*}
satisfy the conditions of Theorem \ref{thm:CNOTtransversality}. Thus these CSS codes are transversal. 
This example shows that one can find codes with different parameters and structures to fit particular error model in neighboring Stations of a quantum network. For instance one code can be better protected against $X$ errors and another code against $Z$ errors. Moreover, nonidentical code allow correcting drastically better correlated errors in the neighboring stations compared to using an identical codes in both stations. Detailed research on this will be presented in future works. 
\end{example}

Let us consider now the CZ transversality. In the following theorem, we define the conditions for CSS codes being CZ-transversal. 
\begin{theorem}\label{Lem:CZtrans}
    Codes $\text{CSS}\left( \mathcal{C}_1, \mathcal{C}_2\right)$ and $\text{CSS}\left( \mathcal{C}_3, \mathcal{C}_4\right)$ are CZ transversal (either of them can be control or target code) iff for any $\bm{\psi}^A, 
    \bm{\psi}^B \in \mathbb{F}_2^k$ we have 
    \begin{align}
        \mathbf{x}^A \mathbf{z}^T + \mathbf{y}\left( \mathbf{x}^B + \mathbf{z}\right)^T  = 0,& \qquad \forall \, \mathbf{y}\in \mathcal{C}_2^{\perp}, \, \mathbf{z}\in \mathcal{C}_4^{\perp}\label{Eq:CZcond1}, \\
       & \qquad \mathbf{x}^A =\bm{\psi}^A \mathbf{A},\nonumber \\
       & \qquad  \mathbf{x}^B=\bm{\psi}^B  \mathbf{B},\nonumber \\
    \mathbf{A B}^T = \mathbf{I}. \label{Eq:CZcond2}
    \end{align}
\end{theorem}
\begin{proof}
According to (\ref{eq:CZ as sum}), if we apply CZ operations to the $k$ pairs of logical qubits in the states $|\bm{\psi}^A\rangle$ and $|\bm{\psi}^B\rangle$, we get the state 
\begin{equation}\label{Eq:LogCZDef}
     (-1)^{\bm{\psi}^A (\bm{\psi}^B)^T} |\bm{\psi}^A \rangle_L \otimes |\bm{\psi}^B \rangle_L. 
\end{equation}
At the same time, if we apply CZ gates to the $n$ pairs of code qubits, we get
\begin{align}
	&\text{CZ} (|\bm{\psi}^A \rangle_L \otimes |\bm{\psi}^B \rangle_L) \nonumber \\
 & = \alpha \sum_{\mathbf{y} \in C_2^{\perp}, \mathbf{z} \in C_4^{\perp}}{(-1)^{\left(\mathbf{x}^A + \mathbf{y}\right) \left(\mathbf{x}^B + \mathbf{z}\right)^T} | \mathbf{x}^A + \mathbf{y} \rangle \otimes  | \mathbf{x}^B + \mathbf{z} \rangle } \nonumber \\ 
	& =  \alpha \beta \sum_{\mathbf{y} \in C_2^{\perp}, \mathbf{z} \in C_4^{\perp}}{ (-1)^{ \mathbf{x}^A \mathbf{z}^T + \mathbf{y}\left( \mathbf{x}^B + \mathbf{z}\right)^T} | \mathbf{x}^A + \mathbf{y} \rangle \otimes  | \mathbf{x}^B + \mathbf{z} \rangle } \label{Eq:FinCZ}
\end{align}
where $\alpha = \frac{1}{\sqrt{\left\lvert C_2^{\perp} \right\rvert \left\lvert C_4^{\perp} \right\rvert}}$ and $\beta = (-1)^{\mathbf{x}^A \left(\mathbf{x}^B \right)^T}$. For having a transversal CZ gate, \eqref{Eq:LogCZDef} should be equal to \eqref{Eq:FinCZ}. Thus, from (\ref{Eq:LogCZDef}), (\ref{Eq:LogQuSA}), and (\ref{Eq:LogQuSB}), we get that 
\eqref{Eq:CZcond1} must hold.  Further 
$$
\beta = (-1)^{\mathbf{x}^A \left(\mathbf{x}^B \right)^T} = (-1)^{\bm{\psi}^A \mathbf{A} \mathbf{B}^T \left( \bm{\psi}^B\right)^T}
$$
must be equal to $(-1)^{\bm{\psi}^A  \left( \bm{\psi}^B\right)^T} $, 
for all $\bm{\psi}^A$ and $\bm{\psi}^B$. This is possible iff $\mathbf{A B}^T = \mathbf{I}$. 
\end{proof}
This theorem allows to formulate the following sufficient conditions  for codes CZ-transversality. 
\begin{corollary}\label{cor:CZ}
It is sufficient for codes $\text{CSS}\left( \mathcal{C}_1, \mathcal{C}_2\right)$ and $\text{CSS}\left( \mathcal{C}_3, \mathcal{C}_4\right)$ to satisfy the following conditions in order to be CZ-transversal  (either of them can be control or target code):  
    \begin{equation} \label{Eq:CZsuffcond1}
        \mathcal{C}_1 / \mathcal{C}_2^{\perp} \cong \mathcal{A}_1, \quad  \mathcal{C}_3 \subseteq \mathcal{C}_2, \quad \mathbf{A B}^T = \mathbf{I},
    \end{equation}
    or
    \begin{equation}\label{Eq:CZsuffcond2}
        \mathcal{C}_1  \subseteq \mathcal{C}_4^\perp, \quad  \mathcal{C}_3 / \mathcal{C}_4^\perp \cong \mathcal{B}_1, \quad \mathbf{A B}^T = \mathbf{I},
    \end{equation}
    where $\mathcal{A}_1 \subset \mathcal{C}_4$ and $\mathcal{B}_1 \subset \mathcal{C}_2$ are group of vectors $\mathbf{c} \in \mathbb{F}_2^n$. 
\end{corollary}

\begin{proof}
In order to satisfy \eqref{Eq:CZcond1}, it is enough that one of the following conditions holds 
\begin{itemize}
    \item $\mathbf{x}^A \mathbf{z}^T = 0$ and $\mathbf{y} \left( \mathbf{x}^B+\mathbf{z}\right)^T = 0$: this is the case if $\mathcal{C}_1 / \mathcal{C}_2^{\perp} \cong \mathcal{A}_1 \subseteq \mathcal{C}_4$, and  $\mathcal{C}_3 \subseteq \mathcal{C}_2$;
    \item $\mathbf{x}^A \mathbf{z}^T = 1$ and $\mathbf{y} \left( \mathbf{x}^B+\mathbf{z}\right)^T = 1$: this could not happen since $\mathbf{z}$ belongs to a linear code and therefore could be $(0,\ldots,0)$; 
    \item $\left( \mathbf{x}^A+ \mathbf{y}\right) \mathbf{z}^T = 0$ and $\mathbf{y}\left( \mathbf{x}^B \right)^T = 0$: this is the case  if $\mathcal{C}_1  \subseteq \mathcal{C}_4^\perp$, and $  \mathcal{C}_3 / \mathcal{C}_4^\perp \cong \mathcal{B}_1 \subseteq \mathcal{C}_2$; 
    \item $\left( \mathbf{x}^A+ \mathbf{y}\right) \mathbf{z}^T = 1$ and $\mathbf{y}\left( \mathbf{x}^B \right)^T = 1$: this could not happen since $\mathbf{z}$ belongs to a linear code.
\end{itemize}
The fact that $\mathbf{AB}^T = \mathbf{I}$ should be satisfied as well completes the proof.
\end{proof}
Below we provide an example of CSS codes that  CZ-transversal. 
\begin{example}\label{Examp:CZtr}
In \cite{DaweConfPaper23} we proposed the mirrored CSS codes that are defined by following generators
\begin{equation}\label{eq:mirroredGenerators}
\begin{bmatrix}
 \mathbf{G_2^\perp} & \mathbf{0} \\
 \mathbf{0} & \mathbf{G_1^\perp}
\end{bmatrix} \mbox{ and }
\begin{bmatrix}
    \mathbf{G_4^\perp} & \mathbf{0} \\
    \mathbf{0} & \mathbf{G_3^\perp}
\end{bmatrix}
=
\begin{bmatrix}
    \mathbf{G_1^\perp} & \mathbf{0}\\
    \mathbf{0} & \mathbf{G_2^\perp}
\end{bmatrix}. 
\end{equation}
We further proved that these codes are CZ-transversal. Below we consider an example of such codes and prove that they are CZ-transversal using Corollary \ref{cor:CZ}. We consider mirrored CSS codes defined by 
$$
\mathbf{G_1^\perp} =\begin{bmatrix}
    1 & 1 & 0 & 0 & 1 & 0 & 0\\
    1 & 1 & 1 & 0 & 0 & 1 & 0\\
    1 & 1 & 1 & 0 & 0 & 0 & 1 
\end{bmatrix},
$$ 
and
$$
\mathbf{G_2^\perp} =\begin{bmatrix}
    1 & 1 & 0 & 0 & 0 & 0 & 0\\
    0 & 1 & 0 & 1 & 1 & 1 & 1 
\end{bmatrix}. 
$$
After simple manipulations we find that 
\begin{align*}
    \mathbf{G}_4= \mathbf{G}_1 =
    \begin{bmatrix}
        \mathbf{G_2^\perp}\\
        \mathbf{A}
    \end{bmatrix}
    & = 
    \left[ 
    \begin{array}{ccccccc}
        1&1&0&0&0&0&0 \\
        0&1&0&1&1&1&1\\
        \hdashline
        0&0&1&1&0&1&1 \\
        1&0&1&1&1&0&0
    \end{array} \right], \\
    \mathbf{G}_2=\mathbf{G}_3 =
    \begin{bmatrix}
        \mathbf{G_4^\perp}\\
        \mathbf{B}
    \end{bmatrix}
    & = 
    \left[ 
    \begin{array}{ccccccc}
        0&0&0&0&1&1&0 \\
        0&0&0&0&1&0&1 \\
        1&1&0&0&0&1&0 \\
        \hdashline
        0&1&1&1&0&0&1 \\
        0&1&1&0&0&0&1
    \end{array} \right].
\end{align*}
We see that $\mathcal{C}_1 / \mathcal{C}_2^\perp \subset \mathcal{C}_4$, $\mathcal{C}_3 = \mathcal{C}_2$,
and $\mathbf{A}\mathbf{B}^T = I_2$.  Thus the conditions 
(\ref{Eq:CZsuffcond1}) are satisfied and the mirrored  CSS codes are indeed CZ-transversal. 
\end{example}
This examples prompts us to think that we can use Corollary \ref{cor:CZ} for giving an alternative proof (to the proof of  CZ-transversality of the mirrored CSS codes.  This is indeed the case. 
\begin{theorem}
    CSS codes with generators defined in (\ref{eq:mirroredGenerators}) are CZ-transversal. 
\end{theorem}
\begin{proof}
The only nontrivial part to prove is to show that matrices 
$\mathbf{A}$ and $\mathbf{B}$ can be chosen so that $\mathbf{A}\mathbf{B}^T = I$. 

We first show that $\mathbf{U}=\mathbf{A}\mathbf{B}^T$ has the full rank. 

The $(i,j)$-th entry of $\mathbf{U}$ is  
$\mathbf{u}_{i,j} = \mathbf{a}_i \mathbf{b}_j^T$, where $\mathbf{a}_i$ and $\mathbf{b}_j$ are $i$th and $j$th row of $\mathbf{A}$ and $\mathbf{B}$, respectively. Assuming that $\mathbf{U}$ is not of full rank, we have that its columns are linear dependant, i.e., $\mathbf{a}_i\left(\mathbf{b}_1+...+\mathbf{b}_k\right)^T = 0$, which means that $\mathbf{a}_i \in \mathcal{B}$. 
It is easy to see that for mirrored CSS codes we have 
$$
\mathbf{G_1} =\begin{bmatrix}
    \mathbf{G_3^\perp}\\
    \mathbf{A}
\end{bmatrix},\ G_3=\begin{bmatrix}
    \mathbf{G_1^\perp}\\
    \mathbf{B}
\end{bmatrix}. 
$$
From this it follows that $\mathcal{R}_{\mathbf{G}_3^\perp} = \mathcal{R}_{\mathbf{G}_1} \cap \mathcal{R}_{\mathbf{B}^\perp}$, where $\mathcal{R}_{\mathbf{G}}$ is the row space of matrix $\mathbf{G}$. Thus, we can conclude that $\mathbf{a}_i \in \mathbf{G}_3^\perp$, which means that $\mathbf{G}_1$ is not of full rank, and therefore this is a contradiction. 

From the fact that $\mathbf{U}$ has the full rank it follows that using the Gaussian elimination we can find $\mathbf{W}$ such that $\mathbf{W}\mathbf{U} = \mathbf{W AB}^T = \mathbf{A}' \mathbf{B}^T = \mathbf{I}$. 
\end{proof}

\section{Relation of Encoding and Transversality}\label{Sec:MappingEffect} 
As we showed in the previous Sections the matrices $\mathbf{A}$ and $\mathbf{B}$ introduced  in (\ref{Eq:LogQuSA}) and (\ref{Eq:LogQuSB}) define the transversality of the corresponding CSS codes. 

It is worth noting that  $\mathbf{A}$ and $\mathbf{B}$ also define the encoding mapping from logical qubits to physical qubits. Let $\psi_i^A = (0, \ldots, 0, 1, 0,\ldots, 0)$, where $1$ is located at the $i$-th position. Then, according to \eqref{Eq:LogQuSA}, we have 
$$
|\psi_i^A\rangle =  \frac{1}{\sqrt{\mathcal{C}_2^\perp}}\sum_{\mathbf{y}\in \mathcal{C}_2^\perp}{\vert\mathbf{a}_i+ \mathbf{y} \rangle}. 
$$
For the same reason we have 
$$
 \vert 0,0, ..., 0 \rangle_L  = \frac{1}{\sqrt{\mathcal{C}_2^\perp}} \sum_{\mathbf{y}\in \mathcal{C}_2^\perp}{\vert \mathbf{y} \rangle}. 
$$
If we denote by $\mathbf{X}_L^i \in \mathbb{C}^{2^k}$ the operator that acts by $\mathbf{X}$ on qubit $i$, and does not rotate other qubits, then from the above two equations we have 
\begin{equation}\label{eq:Xlogical}
    (\mathbf{X}_L^i \vert 0,0,..., 0 \rangle)_L =  \frac{1}{\sqrt{\mathcal{C}_2^\perp}} \sum_{\mathbf{y}\in \mathcal{C}_2^\perp}{\vert \mathbf{a}_i + \mathbf{y} \rangle}. 
\end{equation}

This means that encoded logical gate $\mathbf{X}_L^i$
corresponds to application of 
$\mathbf{a}_i$   to physical qubits. 

From Theorem \ref{thm:CNOTtransversality} we know that in order 
for CSS codes  used in Station A and Station B to be CNOT-transversal it is necessary that $\mathbf{A}=\mathbf{B}$.  
Putting this together with (\ref{eq:Xlogical}), we conclude that for CNOT-transversality it is necessary that implementations of encoded logical $X$ gates for these codes (or equivalently encoding in Stations A and B)  be identical, despite that the codes themselves are different.

Let us now consider encoded CZ-transversality.
First, we
would like to recall that it was shown in [12], [13] that any self-orthogonal CSS code, that is CSS(C,C) code, is CZ-transversal.
 This means that one can use any encoding and still have CZ-transversality between any two code vectors   However, 
from Theorem \ref{thm:CNOTtransversality} and Corollary \ref{cor:CZ} it follows that it is not the case for code vectors from two different CSS codes, which could be used, for example, in Stations A and B. Indeed in this case the encodings must be such that to guarantee the property $\mathbf{A}\mathbf{B}^T\neq\mathbf{I}$.

One more remark is that self-orthogonal CSS codes, that is $CSS(\mathcal{C},\mathcal{C})$ codes, are a special case of the  mirrored structure defined in (\ref{eq:mirroredGenerators}). Thus our proof of CZ-transversality of the mirrored CSS codes gives an alternative proof that self-orthogonal CSS codes are CZ-transversal.  

As a conclusion of this section we would like to point out that 
the use of the same self-orthogonal CSS code in Stations A and B gives more flexibility in terms of encoding operations in Stations A and B. However, using nonidentical CZ-transversal CSS codes in neighboring 
 stations provides in order of magnitude improvement in the fidelity compared with the case of using the same codes \cite{DaweConfPaper23}.

\section{Conclusion}\label{Sec:Concl}

Since different stations in the quantum repeaters may experience different errors, in this paper we consider the second generation QRs with different CSS codes in nearby stations. In the considered setup one important concern should be answered is the transversality of CNOT or CZ gates. So, we first investigated the transversality of the non-local CNOT gates and found a less restrictive constraints  than the case of having the same CSS codes in the nearby station. Also, we found out sufficient condition for having non-local transversal CZ gate in the nearby stations. Finally, we showed the importance of the mapping in the transversality investigation. We observed that self-orthogonal codes with arbitrary mapping would not lead to transversal CZ gate. Possible future direction would be investigating a gate from higher level of the Clifford hierarchy such as T-gate to examine the possibility of having universal transversal gate set using different codes in nearby stations.

\bibliographystyle{IEEEtran}
{\footnotesize \bibliography{IEEEabrv,ASSref}
}
\end{document}